# To Be, Or Not To Be?: Regulating Impossible AI in the United States

**Maanas Kumar Sharma[1]**

[1] Department of Electrical Engineering and Computer Science, Massachusetts Institute of Technology, Cambridge, MA 02139

**Abstract**

Many AI systems are deployed even when they do not work. Some AI will simply never be able to perform the task it claims to perform. We call such systems Impossible AI. This paper seeks to provide an integrated introduction to Impossible AI in the United States and guide advocates, both technical and policy, to push forward regulation of Impossible AI in the U.S. The paper tracks three examples of Impossible AI through their development, deployment, criticism, and government regulation (or lack thereof). We combine this with an analysis of the fundamental barriers in the way of current calls for Impossible AI regulation and then offer areas and directions in which to focus advocacy. In particular, we advance a functionality-first approach that centers the fundamental impossibility of these systems and caution against criti-hype. This work is part of a broader shift in the community to focus on validity challenges to AI, the decision not to deploy technical systems, and connecting technical work with advocacy.

## 1. Introduction

Over the last ten years, AI fairness and ethics has become more prominent in academia as well as in broader society (Pessach and Schmueli 2022). However, there are still significant dangers in our approach to AI (Strickland 2022; Rainie, Anderson, and Vogels 2021). A core piece of the puzzle is to choose to not deploy AI systems in certain situations, though this rarely happens (Barocas et al. 2020; Strickland 2022). We focus on one category of systems that are particularly ripe for "no-go" decisions: *Impossible AI*. We define Impossible AI as systems that purport to perform tasks that are fundamentally unachievable, that is where "no specific AI developed for the task can ever possibly work" (Raji et al. 2022).[1] Common examples of Impossible AI include predicting the emotion or personality traits, like criminality, of people (Fussell 2020; Wu and Zhang 2016), identifying sexuality from facial images (Rotenberg et al. 2024b; Agüera y Arcas, Todorov, and Mitchell 2018), and predicting crime location or perpetrators (Ferguson 2016).[2]

These tasks are considered impossible for slightly different reasons. For example, the very construct of "criminality" is invalid as a human trait (Coalition for Critical Technology 2020; Agüera y Arcas, Todorov, and Mitchell 2017). On the other hand, though sexuality is a veritable aspect of a person's identity, it is impossible to map any subjective social construct to physiological features—though many have tried for centuries as part of the pseudoscience of *physiognomy* (Agüera y Arcas, Todorov, and Mitchell 2017). In all cases, however, the core task of Impossible AI can itself not be validly created.

Examples of Impossible AI are commonly found among academia, private industry, and government across the world. Many academic researchers develop and defend physiognomic AI like those that predict sexuality or criminality (Stark and Hutson 2021; Ford 2022). There are thousands of articles published in affective computing, the super-field that contains emotion recognition AI, and companies capitalize on this technology (Pei et al. 2024). The global artificial intelligence emotion recognition market is predicted to exceed $70 Billion by 2030 (ResearchAndMarkets 2023). In the U.S., major companies like Amazon and Microsoft offer software that claim to predict emotions (Amazon Rekognition n.d.; PatrickFarley and eric-urban 2023). Some go even further, like the Israeli company Faception, which claims to predict whether a person is a terrorist, untrustworthy, or smart based just on their facial image (Faception n.d.). With lofty promises in sensitive settings like public safety—Faception says "It's objective like blood testing"—governments often fund and purchase these technologies as well (Faception n.d.). In 2018, the European Union (EU) began to use an AI system at the border that purported to use "the micro-expressions of travelers to figure out if the interviewee is lying" (Boffey 2018). The Chinese government relies on a web of technology firms, small specialized ones as well as large multinational corporations

---

[1] We largely borrow from Raji et al.'s introduction of "impossible tasks" to the literature, but instead focus on systems that attempt to perform those tasks. To our knowledge, we are the first to define and consider Impossible AI as a category. Examples of what we call Impossible AI have been called pseudo-scientific or invalid by some (Rotenberg et al. 2024b; Coston et al. 2023).

[2] We recognize that these might be contentious classifications (see section 3). This article does not intend to litigate whether certain examples should be called Impossible AI or not; we use those examples that are commonly found in the literature.

like Lenovo, to monitor the emotional states in settings ranging from schools to Uyghur surveillance (Asher-Schapiro 2021; Wakefield 2021; ARTICLE 19 2021).

Many, especially civil society advocates, have rallied against these technologies. A striking example is the fight to prohibit certain Impossible AI systems in the EU AI Act. Initial "red lines" for unacceptable uses in the AI Act did not include any Impossible AI systems. For example, the initial draft only mandated notification to users and additional data privacy for biometric emotion recognition technology (Veale and Borgesius 2021). Very quickly, advocates argued that these restrictions "risk[ed] legitimising a practice [emotion recognition AI] with little-to-no scientific basis and potentially unjust societal consequences" (Veale and Borgesius 2021). A large coalition of civil society groups stated that emotion recognition is prima facie unachievable and pseudoscientific (Access Now et al. 2022b). They argued that emotion recognition relied on "Basic Emotion Theory," which claims that there is a uniform link between physical expressions and inner emotional state (Pei et al. 2024; Marda and Jakubowska 2023). Basic Emotion Theory has been repeatedly repudiated by modern science: in its existence, in its reliability, and in its uniformity across cultures (Barrett et al. 2019; Korte 2020; Krys et al. 2016). Civil society groups also rallied for prohibitions on other examples of Impossible AI: gender and sexuality predictions and predictive policing (All Out n.d.; EDRi et al. 2023).

The finalized AI Act reflected the influence of this advocacy, adding new red-line bans to its 2021 instantiation, but was significantly pared down over time due to political compromise (Jakubowska 2023; Jakubowska et al. 2024). More specifically, it prohibited emotion detection systems in the work and education contexts but exempted uses to "detect, prevent, and investigate criminal offenses" (European Parliament 2024; Borak 2024). The final Act also banned using AI on biometric information to predict "political opinions, trade union membership, religious or philosophical beliefs, race, sex life or sexual orientation" (European Parliament 2024). However, it ignored gender identity predictions and place-based predictive policing and contained relatively open exemptions for law enforcement and government agencies on all prohibitions (Jakubowska et al. 2024).

In the US, however, there is significantly less progress on regulating any form of Impossible AI. No major policy from the Executive branch or Congress has addressed any examples of Impossible AI (Thierer 2024c), though civil society groups are also concerned in the U.S.: the Center for AI and Digital Policy called for bans on all "pseudoscientific and human-rights impacting systems" and the Brookings Institution recommended a ban on all affective computing in federal law enforcement (Rotenberg et al. 2023b; Engler 2021). Why is there no meaningful action in the United States on Impossible AI? What are feasible paths that those who wish to regulate Impossible AI to accomplish such goals?[3]

It is these questions that this article turns to. We focus on three common examples that could be classified as Impossible AI: emotion or personality analysis, predictive policing, and gender and sexuality prediction (Raji et al. 2022). This section provided a primer on Impossible AI and a broad overview of its development and advocacy against it across the world. In Section 2, we explore previous work related to this paper and introduce the three running Impossible AI examples we consider. Section 3 provides a first of its kind catalog of responses from the U.S. federal government—Executive and Legislative—to advocacy surrounding the three examples. In Section 4, we analyze factors of the current policy and advocacy environment that may serve as fundamental blocks to enacting Impossible AI regulation. In particular, we discuss how current advocates use of criti-hype is dangerous. Finally, in Section 5, we discuss paths forward that are more likely to bear fruit in Impossible AI regulation. Section 6 concludes.

## 2. Related Work

### Refusal to Design and Deploy

First, this work draws on discussions of when to not create or use certain technologies. This line of thinking has existed for a long time and in different contexts (Baumer and Silberman 2011; Tierney 2019) but has been of renewed interest in the machine learning community recently (Barocas et al. 2020). Recently, important work has demonstrated how AI systems often do not work (Broussard 2019; Narayanan and Kapoor 2024) and exposed the "fallacy of AI functionality"—the common assumption of AI systems to have some sort of innate capability/expertise, and just need to be "fixed" to be more fair, ethical, safe, trustworthy, etc.—and has instead grounded deployment decisions in real discussion of their validity or functionality (Raji et al. 2022; Coston et al. 2023). Wang et al. (2024) offer a more sweeping critique by considering a number of predictive optimization systems—ML models that predict individuals' future outcomes for decision-making—and conclude that predictive optimization is so structurally flawed in current deployments that "predictive optimization should be considered presumptively illegitimate" and developers and deployers must justify that their model specifically overcomes these burdens (Wang, Kapoor, et al. 2024). Our work is situated within this key idea of refusing rather than reforming AI systems but is oriented towards Impossible AI specifically.

---

[3] It is not necessarily our position that all Impossible AI should be completely prohibited; defending a specific policy is not the purpose of this article. We do recognition combatting Impossible AI as a generally important direction, based on previous work, and seek to advance understanding of the technical and policy landscape of this front. More fundamentally, we believe that advancing scholarship on specific beachhead opportunities in AI regulation in the United States is net beneficial for ensuring the fair, ethical, and safe use of AI.

## Impossible AI

Many systems could be considered Impossible AI. We will not argue for or against certain categorizations, but rather provisionally accept examples used in the literature and commonly described as "pseudoscientific" or "invalid" by advocates (Raji et al. 2022; Access Now et al. 2022b; All Out n.d.; Rotenberg et al. 2023b). We will specifically focus on predicting emotion or personality (hereinforth *emotion AI*), predictive policing, and gender or sexuality prediction, for future discussion. Each of these fields are considered in varying ways in the literature, and we catalog them below.

### A) Emotion AI

There is a wide spectrum of perspectives on emotion AI in the CS community. Significant portions of the community unflinchingly work on advancing the technical "ability" to recognize emotions through machine learning (Keesing et al. 2023). Other work is critical of emotion AI but believes in its fundamental validity. For instance, Kim et al. points out disparities on different age groups, but argues for "inclusive, intersectional algorithmic developmental practices" to achieve the "potential for facial emotion recognition" (2021). Bryant et al. instead advocates for new data collection practices to address the ambiguous perception of facial expression (2022). Other work doubts the basic validity of emotion AI, but primarily advances more pragmatic critiques of the real and often-used technology (Boyd and Andalibi 2023; Diberardino and Stark 2023). Finally, Agüera y Arcas et al. primarily critique the physiognomic underpinnings of emotion AI and other systems (2017).

### B) Predictive Policing

There has been much work done on predictive policing. Academics, civil society, and media have all worked on establishing the bias of predictive policing (Selbst 2017; Heaven 2020; Rotenberg et al. 2024a; Camilleri et al. 2023). A more recent focus has also been the inefficacy of these programs (Sankin and Mattu 2023). Work in the computer science community has often focused on understanding and addressing the feedback loops that plague predictive policing (Ensign et al. 2018; Akpinar, De-Arteaga, and Chouldechova 2021; Pagan et al. 2023; Biswas et al. 2023). Others have advanced more community-centered, critical theory-based approaches to race and predictive policing (Hanna et al. 2019; Moorosi et al. 2023; Jegede et al. 2023). Community movements like the mathematician's boycott of police work also reflect these perspectives (Aougab et al. 2020).

### C) Gender and Sexuality Prediction

The literature has also been critical of technology that seeks to infer gender or sexuality.[4] Assessing gender through facial images or other data is common throughout society, from standard computer vision models to fitness trackers (Ovalle, Liang, and Boyd 2023; Scheuerman et al. 2020). However, this process often assumes a binary, static, and biologically determined sex and/or gender identity— which is not scientifically accurate (Fuentes 2022; King 2022; All Out n.d.).

In the literature, Scheuerman et al. describes how dataset annotation assigns an indisputable gender that is propagated through machine learning pipelines (2020). Hamidi et al. offers community-oriented perspectives on automatic gender recognition and their profound harms through interviews with transgender people (2018). Ovalle et al. consider these same considerations in new technologies like smartphone biometric data and the failure of AI development and governmental protections to protect the queer community (2023). Finally, gender AI systems have intersectional effects with racial and other discrimination in AI and society (Buolamwini and Gebru 2018; Few 2007).

Sexuality detection is less widespread than gender detection, but academics do attempt to use deep learning on facial, biometric, or medical data to predict sexual orientation (Wang and Kosinski 2018; Ziogas et al. 2023). However, these are condemned more often with a belief that such a prediction task is fundamentally impossible (Cockerell 2023; Agüera y Arcas, Todorov, and Mitchell 2018).

## AI Governance

The AI community has increasingly considered AI governance as well, both for predictive and generative AI systems (Lucaj, van der Smagt, and Benbouzid 2023). As a starting point, we note work that establishes the innovation-driven, hands-off approach to regulating information technology in the United States (Bradford 2023; Frank 2023). More specifically, much work recently has focused on designing and implementing algorithmic auditing (Raji, Costanza-Chock, and Buolamwini 2024; Birhane et al. 2024; Casper et al. 2024). Other work discusses AI governance within organizations that develop or deploy AI (Gill et al. 2022). Spurred by newer, highly capable generative models, discussion has also flurried around proposals like disclosure, registration, and licensing (Guha et al. 2024; Stanford Human-Centered Artificial Intelligence n.d.). Finally, recent work has thoughtfully considered the range of preexisting regulatory mechanisms that could be applied to regulate AI

---

[4] However, the academic community as a whole has much to work towards trans- and queer-inclusive research (Keyes 2018).

in the United States (Raji et al. 2022; Bondi-Kelly et al. 2023).

## 3. The Current State of Impossible AI Regulation

Although the U.S. is known for a lax regulatory environment for technological innovation (Bradford 2023; Kang 2022), the federal government has taken up the issue of AI with fervor in all three branches. Congress has held many hearings to understand and discuss AI, both in general and in specific areas, and a number of bills, many of them bipartisan, have been proposed, though none have progressed far (Rotenberg 2023; Poinski 2024). The White House has advanced AI policy through voluntary commitments from large AI corporations, publishing the Blueprint for an AI Bill of Rights, and the Executive Order on the Safe, Security, and Trustworthy Development and Use of Artificial Intelligence (AI Executive Order) (Harvard Law Review 2024; OSTP 2023; The White House 2023). Executive Agencies—especially the National Institute of Standards and Technology (NIST), the National Science Foundation (NSF), and the Federal Trade Commission (FTC), but many more as well—are also playing a large part in regulating AI using their existing powers (Rotenberg 2023; The White House 2024). Legal claims against AI—both predictive and generative—have also cropped up in federal courts across the country (Albarazi 2024; Cusenza 2024b). Court cases have yet to significantly alter the legal regime around AI, but could potentially be significant, especially with legislative gridlock (Rotenberg 2023; Albarazi 2024). However, these actions have not addressed Impossible AI, either as a category or through specific examples, in any meaningful way. Importantly, this does not necessarily imply a lack of movement from civil society. In this section, we will catalog both the current AI governance affecting, and the movements against, our three examples of Impossible AI in the United States.

### A) Emotion AI

Public opposition in the U.S. against emotion analysis sprouted in early 2019, largely based on concerns about its equity, alongside revelations of enormous bias in other facial analysis AI (Rhue 2019; Heckman 2020). However, advocates soon began to take more fundamental challenges to the technology. In 2019, The AI Now Institute characterized affect recognition as "built on markedly shaky foundations … and at worst entirely lack[ing] validity," but only recommended prohibiting its use in high-stakes decision-making, not research or development as a whole (Crawford et al. 2019). Meredith Whittaker, a co-founder of the AI Now Institute, testified in a January 2020 hearing before the House Oversight Committee that affect recognition is "not supported by scientific consensus and recalls discredited pseudoscience of the past" (U.S. Government Publishing Office 2020). Since then, the Brookings Institution recommended a federal ban on affective computing in law enforcement, also due to "unreliable" behavior and little "scientific basis," but did not recommend an outright ban due to "plausible valuable contributions … that warrant further research" (Engler 2021). However, the Center for AI and Digital Policy (CAIDP) has recommended an outright ban on sentiment detection and analysis (Rotenberg et al. 2023b; Rotenberg et al. 2024b). Activists have also responded to specific instances of these technologies. For instance, in 2022, a group of more than 25 organizations sent an open letter to Zoom urging it to stop exploring emotion detection technology (Access Now et al. 2022a) and CAIDP submitted a complaint to the FTC regarding Zoom's use of AI (Rotenberg et al. 2023a).

The federal government's response to these calls has been enormously lacking. We only found one substantive Congressional discussion of emotion AI after that 2020 hearing[5] (U.S. Government Publishing Office 2023). In June 2023, Senator Ron Wyden (D-OR) publicly supported the EU's proposed ban of emotion detection AI and called the technology "bunk science" (Wyden 2023). However, NIST defended some emotion AI as "emerging experimental" work and characterized errors in the systems as "spurious correlations" (Schwartz et al. 2022). As such, there is literal policy momentum for regulating emotion; momentum, including by Wyden, leans more towards the general case of regulating "flawed automated decision systems" (Wyden 2023). Additional discussion of emotion AI is only in the context of fairness and bias, which risks legitimizing the inherent premise of the technology (Venkatasubramanian 2023). President Biden's AI Executive Order, one of longest Executive Orders ever, does not mention emotion detection AI at all.[6] A few sub-federal governments have taken action for particular examples of emotion AI, like in predicting traits of job fit in employment decisions (Heilweil 2020; Chen and Hao 2020). However, these laws are few and have extremely limited protections, like notice and bias audits (Heilweil 2020; Kestenbaum 2023; HRDrive 2023).

### B) Predictive Policing

---

[5] A hearing before the House Committee on Science, Space, and Technology in June 2019 did discuss similar issues, but occurred before (U.S. Government Publishing Office 2019).

[6] The closest reference is a direction for the Architectural and Transportation Barriers Compliance Board to "issue technical assistance and recommendations on the risks and benefits of AI in using biometric data as an input" for people with disabilities, "including gaze direction, eye tracking, gait analysis, and hand motions." (The White House 2023).

Community opposition to predictive policing has largely focused on place-based predictive policing, the most commonly used technology, versus incident-based or individual-based predictive policing, which is less commonly used and more experimental (Lau 2020; Ferguson 2016). Predictive policing is often characterized as a "self-fulfilling prophecy" and discriminatory for its reliance on biased historical data (Haskins 2019; NAACP n.d.). Academics, media, and civil society have all challenged predictive policing through research, advocacy, and legal challenges (Ensign et al. 2018; Saunders, Kunt, and Hollywood 2016; The New York Times 2015; Lau 2020; Brennan Center for Justice 2021). As a result, some police forces, including Los Angeles, New York, and Chicago, shelved certain predictive policing systems (Lau 2020). Some recent advocacy has shifted focus from the biased results to the overall inefficacy of these systems (Sankin and Mattu 2023), calling for bans on predictive policing like those in Oakland and Santa Cruz (Guariglia and Kelley 2023; Johnston 2020; Asher-Schapiro 2020). Even fewer characterize predictive policing as not just ineffective currently, but fundamentally pseudoscientific or invalid (Rotenberg et al. 2024a).

Though some local governments have been quite responsive to these calls, the federal government is less so, though it does still acknowledge predictive policing as a possible harm. A group of Democratic Senators and Representatives penned an open letter to the Justice Department at the beginning of this year asking it to stop federal funding for predictive policing, "unless it is proven not to discriminate and meets standards for effectiveness and accuracy" (Wyden et al. 2024). However, this issue divides Congress. In a hearing on AI in Criminal Investigations and Prosecutions, Republican Senators expressed strong support for AI "modeling where and when crimes happen" (U.S. Senate Committee on the Judiciary 2024).

President Biden's AI Executive Order tasked the Attorney General with creating a report on the use of AI in the criminal justice system, including predictive policing, and how to ensure "fair and impartial justice for all with respect to the use of AI in the criminal justice system" (The White House 2023). However, previous rulemaking on the use of AI by federal agencies is rife with exemptions for law enforcement and mired by transparency and compliance issues (Buolamwini and Friedman 2024; NAIAC Law Enforcement Subcommittee 2024; GAO 2024). Additionally, previous work from the President's National Artificial Intelligence Advisory Committee (NAIAC) has focused on bias and data acquisition while completely ignoring the validity of the prediction task itself (NAIAC 2023; NAIAC-LE 2024).

*C) Gender and Sexuality Prediction*

Although there was a large push to ban AI sexuality and gender inference in the EU (All Out n.d), there is no real corollary for that in the U.S. Some media discussion, for instance after Uber driver verification algorithms kicked transgender drivers off the app, and advocacy rallies against gender and sexuality detection, but the work is few and far in between (Urbi 2018; Samuel 2019; West, Whittaker, and Crawford 2019; Cockerell 2023).

There is also scant government consideration of these criticisms. A witness shared the Uber example in a 2019 House Committee on Science, Space, and Technology hearing, but we did not find any other discussions in Congressional hearings (U.S. Government Publishing Office 2019). The AI Risk Management Framework, NIST's flagship voluntary guidance to organizations developing AI systems, does not mention LGBTQ+ people or gender/sexuality detection even once (NIST 2023). Even in NIST's report specifically on identifying and managing AI bias, though it mentions that sexual orientation inferences are "not scientifically supported," it stays far from a compelling normative claim (Schwartz et al. 2022). Worse, in its discussion of "systemic bias in gender identification," the report largely only discusses racial bias while completely ignoring the task's validity and relegating transgender and gender-nonconforming people to a final cursory sentence on the "lack of awareness about the multiplicity of gender" (Schwartz et al. 2022). Slightly better, in its report examining AI harms, NAIAC wrote that identity verification can "(1) out trans people and cause them gender dysphoria; (2) incorrectly classify gender minorities as security risks; and (3) discriminate against gender minorities trying to enter the U.S. or access essential health, employment, and housing services" (NAIAC 2023). However, NAIAC is only an advisory group to the President, and its report has little legal or even political effect.

## 4. Barriers to Impossible AI Regulation

### Barrier 1: Politics and Procedure

It is well-known that the U.S. has been loath to regulate the internet and new digital technologies, especially when compared to Europe and China (Bradford 2023). In the words of Anu Bradford, "The American regulatory approach centers on protecting free speech, the free internet, and incentives to innovate. It is characterized by its discernible techno-optimism and relentless pursuit of innovation" (2023). Furthermore, history and research suggests that policymakers begin to support regulation only from multiple high-profile failures or harms to individual liberty, rather than social or ethical claims (Jensen 2015; Coglianese 2012; Hansen, McAndrews, and Berkeley 2008; Beens 2020; Stiglitz 2024; Schiff 2023). This ethos will complicate pushes to regulate Impossible AI, as advocates argue that the harms of Impossible AI are unique from other AI harms and ought to be addressed through a more aggressive approach.

Moreover, even when public opinion supports certain regulation, the U.S. has failed to pass it, largely due to the structure of the American political system (Godlasky 2022). Countless examples illustrate how an independently elected executive, a strong bicameral legislature, the committee system and other Congressional procedures all make legislation exceptionally difficult to pass Congress (Scalia 2011; Godlasky 2022; Barrow 2016; Willis and Kane 2018). In contrast, the EU's structure limits public backlash to decision makers, has easier passage standards, defers more to experts, and promotes upward harmonization of regulatory standards across the EU (Bradford 2020) while China's authoritarian party rule renders regulation an easy top-down exercise of power (Chen 2024).

### Barrier 2: Controversial Foundations

A core hurdle in advocating for strict regulation on Impossible AI is the difficulty to argue that the tasks themselves are impossible. Often, these overarching claims rely on philosophical or highly contested arguments. Even in the case of emotion detection, where the science is relatively clear that emotions are not statically nor uniformly expressed or perceived (Barrett et al. 2019; Krys et al. 2016), many—in research, industry, and the general public alike—believe this a hurdle, not a wall (Somers 2019; Telford 2019; Castro 2023). The American ethos stands behind constant "innovation" and is loath to characterize tasks as impossible, which manifests clearly in the response to challenges to emotion AI (Hallonsten 2023; Winner 2018; Telford 2019). For predictive policing, claims about functionality are entangled in political beliefs surrounding crime and policing (Ekins 2016; Malik 2021; Drakulich 2022; Hvistendahl 2021). Gender and sexuality detection are also caught up in personal beliefs that can contradict scientific consensus (Parker, Horowitz, and Brown 2022; Feeney 2022; The National Academies of Sciences, Engineering, and Medicine 2022). Finally, given the benefits these technologies could provide (with the hidden assumption that they work) and the fact that AI policy has thus far prioritized economics and innovation over social and ethical concerns (Schiff 2023; Schiff 2024), the fight to regulate Impossible AI becomes only more difficult (McGrath and Nnamoko 2023).

### Barrier 3: Criti-hype

To close, we will discuss a pernicious approach in AI criticism that we believe harms the future of regulating Impossible AI: *criti-hype*. First introduced by Lee Vinsel, criti-hype is criticism of technologies that both needs and feeds the hype of the technologies it critiques (2021). Indeed, popular attention of AI has significantly hyped the technology as magical and omnipotent, even though they are riddled with functionality concerns (Narayanan and Kapoor 2022; Narayanan and Kapoor 2023a; Narayanan and Kapoor 2023b; Raji et al. 2022). Shortsightedly, many criticisms of AI and calls for its regulation have built themselves on top of these falsehoods and what harms they could cause (De Vynck 2024; Nolan, Maryam, and Kleinman 2024).

Although these inventions may be effective in spurring public opinion and perhaps the government (Tyson and Kikuchi 2023; Schiff 2024), this approach is a problematic way to approach AI regulation, especially for Impossible AI. First, regulating the "wishful worries" of a mythical version of a technology will not protect us from the "actual agonies" of its true existence (Brock 2019; Narayanan and Kapoor 2023b). For instance, Senator Jeff Merkley (D-Ore.) expressed concern about emotion recognition systems because they can help governments hyper-accurately spot dissidents in public places (Merkley 2021). This claim could motivate a limited government action, say, to ban emotion detection use at protests by police. However, this would not protect many other instances of harm from emotion AI, including in hiring, criminal investigations outside of protests, and much more. Worse, pursuing such a path would reinforce the claims of emotion AI manufacturers that their products are in fact capable of assessing emotions, and accelerate their use in those other sectors where people are unprotected from their harms (Chen and Hao 2020). Instead, advocates for Impossible AI regulation should attempt to remain as principled as possible, focusing on actual agonies over wishful worries, and highlighting the impossible functionality of the systems and the harm they will cause if left unchecked. This approach was modeled by advocates for an emotion AI ban in the EU AI Act. Early drafts focused on mass emotional manipulation, a wishful worry, but focused advocacy recentered the act around the actual agonies of emotion AI (Franklin et al. 2022; Access Now et al. 2022b).

## 5. Paths Forward

Curtailing the deployment of Impossible AI in the United States does not necessarily need to come from legislation that explicitly discusses "Impossible AI" in those terms, or even examples of it explicitly. There are a number of other policy avenues that offer ways to stop certain Impossible AI use. We also note that, stemming from Barrier 3, advocacy against Impossible AI should avoid criti-hype. We believe that an overall shift to focus more on the *(non-)functionality* of AI systems in regulation conversations accomplishes both of these goals (Raji et al. 2022). By centering functionality and the validity of AI systems, general policy on AI is more likely to also include protections against Impossible AI, unlike in the current discourse (see section 3) and be more robust than policy developed through criti-hype. In this section, we will describe some policy mechanisms that we believe are likely to bear fruit

in regulating Impossible AI if such efforts are directed there.[7]

### International Principles

Early work in AI policy came in the form of principles to guide AI development, like the OECD AI Principles, G20 AI Guidelines, and Universal Guidelines for AI (CAIDP n.d.). These guidelines are voluntary principles to inform the design, use, and regulation of AI. Though they have limited direct effect, they are influential in lobbying stakeholders, including the Executive branch, state lawmakers, computing societies, and even industry, to adopt certain behaviors (de Souza Dias and Sagoo; CAIDP n.d.). Indeed, these principles do provide a basis to challenge Impossible AI. For instance, the OECD AI Principles, which the U.S. adheres to, state that "AI systems should … function appropriately" and that AI actors should be accountable for their proper functioning (OECD 2024). However, recent updates to the OECD AI Principles have been criticized by some as walking back the strength of language on the right to contest adverse decisions and combating algorithmic bias (CAIDP 2024). Important work should go into upholding the functionality basis in international AI principles and expanding them to explicitly call for "no-go" decisions when AI systems do not or cannot work as intended.

### AI Standards

Other, more direct, forms of voluntary, non-binding guidance come from agencies in the U.S. itself. In particular, the National Institute of Standards and Technology has been tasked with creating a number of standards for safe, trustworthy, and responsible AI (NIST 2024; U.S. Department of Commerce 2024). As explored in section 3, current NIST guidance currently ignores key functionality concerns, both in general and in specific examples. Standards are political artifacts, and therefore can be influenced by external actors (Solow-Niederman 2024). We note that NIST is a small institute, generally new to AI, and draws on outside expertise and collaboration (Press 2023; Rotenberg et al. 2024b). NIST is also a technical organization, so technical advocacy emphasizing functionality and decisions to not deploy could have a large impact even amidst the complex politics of AI.

### Sector-specific Regulators

Many have suggested that sector-specific regulators are well-poised to address the manifested risks of AI sector-by-sector (Raji et al. 2022; Thierer and Brimhall 2024; Kelley 2024). For instance, the Food and Drug Administration has the authority to regulate AI used in healthcare, the Equal Employment Opportunity Commission (EEOC) for employment, and so on for the many regulatory bodies in the U.S. (Bondi-Kelly et al. 2023; Federal Register n.d.).

Many have already begun to consider the impact of AI on their responsibilities (Thierer 2024d), but these too do not adequately realize that AI products are often non-functional, snake oil, or fully impossible (Raji et al. 2022; Narayanan and Kapoor 2024). For instance, in the EEOC's most recent guidance on disparate impact in employment AI systems focuses entirely on the bias of the deployed tool, not the prior question in disparate impact law of whether "the model adequately predicts what it is supposed to predict" (Barocas and Selbst 2016; EEOC 2023; Davisson, Zhou, and Winters 2023).The EEOC's silence on what is already a central question of its work in employment discrimination is worrying, and indicative that much more work is needed to center other sector-specific regulators on the validity of AI systems when regulating their use. Nevertheless, sector-specific regulation is an important way to check against the use of Impossible AI in specific, particularly worrying sectors.

### Consumer Protection

In the U.S., the Federal Trade Commission (FTC) has the broadest jurisdiction for consumer protection as it is charged with regulating "unfair or deceptive acts or practices in or affecting commerce" (FTC 2021a). The commission has broad investigatory powers, can enter into consent decrees with companies (pre-trial settlements), can levy financial penalties and injunctions through courts, and, again as of 2021, can issue rules (FTC 2021a; Raji et al. 2022; FTC 2021b). The FTC's authority lends itself well to regulating Impossible AI: AI products that claim to perform tasks that they will never be able to could easily be argued as deceptive or unfair.[8] Indeed, the FTC has already written warning companies against "claiming [an AI product] can do something beyond the current capability of any AI or automated technology?", though there are no examples of it taking action against a company for false functionality claims yet (Atleson 2023).

The current political reality is also favorable to this avenue. Under the leadership of current Chairwoman Lina Khan, the agency has been incredibly active, including in the tech sector and for AI specifically (Abovyan and Scanlan 2024; Cohen 2024; Khan 2023). The FTC released multiple guidances to companies developing and deploying AI, streamlined the investigation process for AI, opened a Civil Investigative Demand into OpenAI and gathered information on several other generative AI developers (Cohen 2024; FTC 2023; CAIDP n.d.; FTC 2024). Chairwoman Khan proclaimed in *The New York Times* that "We

---

[7] This is not intended to be a comprehensive enumeration. Indeed, many approaches in AI regulation are not included here, including product liability, auditing, and more (Raji et al. 2022; Guha et al. 2024; Rotenberg 2023).

[8] This line of argumentation will remain susceptible to the barrier of proving such a claim as discussed in section 4.

Must Regulate AI." However, some, especially Republicans, have criticized Chairwoman Khan and her expansive actions, so it is unclear how stable this situation is (U.S. House Committee of the Judiciary 2024; Chilson 2024). Additionally, the FTC is not a regulatory silver bullet even in the Khan administration. For instance, it has lagged on its investigation of OpenAI for over a year and did not respond to a complaint against Zoom for its expansion of AI usage, including for emotion detection (CAIDP n.d.; Rotenberg et al. 2023a) Further, the FTC simply does not have jurisdiction over government (including police departments), nonprofits, and some sectors like transportation, banks, and insurance (Gellman 2016).

### The States

In the United States' federal system of government, states and localities are key policy venues (Peterson 1995). States have acknowledged the importance of AI and the gridlock of Congress, have taken active efforts in regulating AI already, and are receptive to consensus building and technical advocacy (Norden and Lerude 2023; Schiff and Schiff 2023). States have introduced legislation ranging from AI research agendas, to combatting discrimination in decision-making systems, to heavy-handed regulatory frameworks like California's SB-1047 (Norden and Lerude 2023; Thierer 2024b).

States are a meaningful venue for multiple reasons. First of all, that regulation protects the people of that state. Also, state innovations can spur other states and the federal government to adopt similar approaches (Dinan 2008). Furthermore, recent state tech policy has also demonstrated that "the early bird gets the worm" as states have copied other states' bills with few changes, so early adoption of functionality-oriented approaches is key (Gedye and Scherer 2024). Finally, regulation in certain states, like California or New York, can have a "Brussels effect"-esque bite on tech companies with important ties to those states (Thierer 2024b; Bipartisan Policy Center 2024).

It is critical to note that state regulation of Impossible AI is not restricted to just legislation. Many states have the above options (and those not discussed) available to them as well! For instance, all states have consumer protection statutes similar to those that empower the FTC, many states have similar bureaucracies, and all states can regulate their own operations (like education, policing, etc.) unlike the federal government (Carter 2009; Edwards 2006; U.S. Department of Education n.d.; Wex Definitions Team 2020).

To reiterate, this list is not intended to be comprehensive. There is still important work to be done in AI regulation broadly, and for Impossible AI specifically, in other areas. For instance, though comprehensive federal AI legislation is unlikely to pass (Thierer 2024a; Whyman 2023), the area is quite bipartisan, and meaningful bills might pass. Work on these efforts should consider how to include Impossible AI either through the overall mechanism design or specific additions for Impossible AI (Guha et al. 2024). Importantly, technological engagement with the courts could be under-appreciated (Hershberger 2022; Rotenberg 2023). Specifically, product liability law could be well-suited for Impossible AI regulation if case law moves in a certain direction (Villasenor 2019; Vasudevan 2023; Pfeiffer 2023; Raji et al. 2022), though there remain barriers to applying product liability to the field (Raji et al. 2022).

## 6. Conclusion and Further Work

Impossible AI is an important subsection of AI systems to consider and regulate. This paper described Impossible AI development, challenges to Impossible AI, government responses to Impossible AI, barriers to regulating Impossible AI, and certain directions to focus advocacy in the United States. We reiterate our connection to previous work basing advocacy in actual agonies versus wishful worries, centering the functionality of AI systems, and deciding not to deploy AI systems. We hope this work will be of use to advocates both of technical and policy backgrounds, as target venues are receptive to both demographics, and both groups could benefit from this analysis of history and guidance for future work. Future work should expand on our policy recommendations as well as more rigorously examine the category of Impossible AI. Is "Impossible AI" the correct terminology? Which systems should be described as Impossible AI? These questions remain open and ripe for discussion.

# Acknowledgments

This manuscript was prepared as part of the course 6.S977: Ethical Machine Learning in Human Deployments at the Massachusetts Institute of Technology, Spring 2024.